\documentclass[useAMS,usenatbib]{mn2e}  
\usepackage{amssymb, amsmath, graphics}
\usepackage{graphicx}  
\usepackage[usenames]{color}  
\usepackage{aas_macros}

\title[{\it Kepler} observations of a roAp star]{{\it Kepler} observations of a  
roAp star:\\$\delta$\,Scuti and $\gamma$\,Doradus pulsations in Ap stars}  
\author[L. A. Balona et al.]{L. A. Balona$^{1}$, M. S. Cunha$^{2}$,
D. W. Kurtz$^{3}$, I. M. Brand\~{a}o$^{2}$, 
\newauthor{M. Gruberbauer$^{4}$, H. Saio$^{5}$, R. {\O}stensen$^{6}$, V.G.~Elkin$^{3}$, W. J. Borucki$^{7}$,}
\newauthor{J. Christensen-Dalsgaard$^{8}$,  H. Kjeldsen$^{8}$, D. G. Koch$^{7}$} \\
\\
$^{1}$South African Astronomical Observatory, P.O. Box 9, Observatory 7935, 
Cape Town, South Africa\\ 
$^{2}$Centro de Astrofisica e Faculdade de Ci\^{e}ncias, Universidade do Porto,
4150-Porto, Portugal.\\
$^{3}$Jeremiah Horrocks Institute of Astrophysics, University of Central 
Lancashire, Preston PR1\,2HE, UK\\ 
$^{4}$Department of Astronomy \& Physics, Saint Mary's University, Halifax, NS B3H 3C3 Canada\\
$^{5}$Astronomical Institute, Graduate School of Science, Tohoku University, Sendai, 980-8578, Japan\\ 
$^{6}$ Instituut voor Sterrenkunde, KULeuven, Celestijnenlaan 200D, 3001 Leuven, 
Belgium\\
$^{7}$NASA Ames Research Center, MS 244-30,Moffett Field, CA 94035, USA\\
$^{8}$Department of Physics and Astronomy, Building 1520, Aarhus University, 8000 Aarhus C, Denmark
} 
\begin{document}  
  
\date{Accepted .... Received ...}  
  
\pagerange{\pageref{firstpage}--\pageref{lastpage}} \pubyear{2010}  
  
\maketitle  
  
\label{firstpage}  
  
\begin{abstract} 
Observations of the A5p star KIC\,8677585 obtained during the {\it Kepler} 10-d 
commissioning run with 1-min time resolution show that it is a roAp star with 
several frequencies with periods near 10\,min. In addition, a low frequency at 
3.142\,d$^{-1}$ is also clearly present. Multiperiodic $\gamma$\,Dor and 
$\delta$\,Sct pulsations, never before seen in any Ap star, are present in {\it 
Kepler} observations of at least three other Ap stars. Since $\gamma$\,Dor 
pulsations are seen in Ap stars, it is likely that the low-frequency in 
KIC\,8677585 is also a $\gamma$\,Dor pulsation. The simultaneous presence of both 
$\gamma$\,Dor and roAp pulsations and the unexpected detection of $\delta$\,Sct 
and $\gamma$\,Dor pulsations in Ap stars present new opportunities and challenges 
for the interpretation of these stars. 
\end{abstract} 
 
 
\section{Introduction} 
 
The {\it Kepler Mission} is designed to detect Earth-like planets around solar-
type stars \citep{Koch2010}. To achieve that goal, {\it Kepler} will continuously 
monitor the brightness of over 150\,000 stars for at least 3.5\,yr in a 105 square 
degree fixed field of view. Photometric results from the 10-d commissioning run 
show that $\mu$mag precision in amplitude can be attained for the brighter 
stars (8--10\,mag) for long-cadence (29.4-min) exposures.  With this level of 
precision interesting pulsational behaviour never seen before is being found 
in many stars \citep{Gilliland2010}. In addition, {\it Kepler} has a small 
allocation for short-cadence (1-min) exposures. In this mode it is possible to 
detect and study the light variations of short-period pulsating stars such as 
solar-like pulsators, $\delta$\,Sct stars, and rapidly-oscillating Ap stars (roAp 
stars). 

The roAp stars, discovered by \citet{Kurtz1982}, are found amongst the coolest 
subgroup of Ap stars, namely, the SrCrEu group (6\,400--10\,000\,K). About 40 
roAp stars are known at present, with temperatures in the range $6400\,{\rm K}
\lesssim T_{\rm eff} \lesssim 8400$\,K, and exhibiting either single or multiperiodic 
pulsations with periods in the range $5.6 - 21$\,min. These oscillations are 
interpreted as acoustic modes of low degree and high radial order (typically $n>$15), 
which are modified at the surface layers by strong, large-scale magnetic fields. 
In that respect they are quite different from the $\delta$~Sct stars which, having 
similar mass and effective temperature, tend to have oscillations with radial order 
not exceeding 4 or 5. The difference in the radial orders of the oscillations found 
in these two classes of pulsators is thought to result from the difference in the 
regions where the modes are excited. While in roAp stars the oscillations are 
believed to be excited in the region of hydrogen ionization 
\citep{Balmforth2001,Cunha2002,Saio2005,Theado2009}, which is partially or fully 
stabilized against convection by the presence of strong magnetic fields, in 
$\delta$~Sct stars the excitation takes place in the region of second helium 
ionization. 
 
Despite the general understanding of the driving of oscillations in roAp stars, 
there are still a number of puzzling questions to be answered. Whereas models of 
hot roAp stars show that driving of pulsations is possible when convection is 
suppressed, this is not the case for cooler roAp stars. For example, 
\citet{Saio2010} find that all roAp-like pulsations are stable in magnetic models 
of HD\,24712 (HR\,1217; DO\,Eri) in which convection is suppressed. A second 
puzzling aspect is that apparently non-oscillating Ap stars called ``noAp stars''
occupy a similar part of the HR~Diagram as the roAp stars. At 
present it is not clear what determines which of these peculiar magnetic stars 
show high radial overtone p-mode pulsation, and which are apparently non-variable. 
This may either be the result of a selective driving mechanism or simply an 
observational bias. It is hoped that the {\it Kepler Mission}, with its ability to 
detect much lower amplitude roAp stars than has previously been possible, will 
illuminate these problems. 
 
The global magnetic field which is present in all Ap stars plays an essential role 
in roAp oscillations, influencing their geometry, frequencies, and energy balance. 
The presence in some roAp stars of frequency multiplets with spacings exactly 
equal to the angular frequency of rotation led \citet{Kurtz1982} to introduce the 
oblique pulsator model, according to which the observed pulsations are 
axisymmetric about the magnetic axis, which is tilted with respect to the 
rotational axis.  This simple picture is challenged by the fact 
that in some stars the peaks lying symmetrically about the central peak in the 
multiplet do not have identical amplitudes. This could potentially be explained by 
the combined effect of rotation and magnetic field on the pulsations, which may 
break the alignment between the pulsation and magnetic axis 
\citep{bigot2002,gough2005}. However, that requires the magnetic and centrifugal 
effects on the oscillation frequencies to be comparable, something that is not 
expected except, possibly, for particular combinations of magnetic field strength 
and oscillation frequencies \citep{cunha2007}. 
 
Most roAp stars do not show the rotational multiplets predicted by the oblique 
pulsator model either because the rotational period is too long (and the 
multiplets unresolved) or because they have amplitudes below the threshold of 
detectability. Many do, however, show multiple frequencies which in some cases can 
be interpreted in terms of modes of alternating spherical harmonic degree, $\ell$, 
and of consecutive radial orders, $n$. 
 
In the case of linear, adiabatic oscillations, in a spherically symmetric star, at 
high radial order the frequencies of modes of the same degree are repeated at 
approximately regular intervals \citep{Tassoul1980}. This interval, defined as the 
frequency difference between modes of the same $\ell$ but successive radial orders 
$n$, $\Delta\nu = \nu_{n\ell} - \nu_{n-1,\ell}$, is called the large separation, 
and is a measure of the mean density of the star. Moreover, in the same asymptotic 
limit of high radial orders, if modes of alternating even and odd $\ell$ are 
present, then the peaks in the frequency spectrum are separated by 
$\simeq\Delta\nu/2$, while if only modes of the same $\ell$ are present, the 
separation between consecutive peaks is $\simeq\Delta\nu$. 
 
In some roAp stars a pattern of nearly equally spaced frequencies is indeed 
observed. If these are interpreted in accordance with the asymptotic relation 
derived by \cite{Tassoul1980}, the average large separation allows the stellar 
parameters to be constrained. In practice, the presence of a magnetic field breaks 
the spherical symmetry of the pulsating star, and perturbs the oscillation 
frequencies away from the values predicted by this asymptotic relation. This has 
been shown theoretically in a number of studies that considered oscillations in 
the presence of a large scale magnetic field \citep{Cunha2000, Saio2004, Saio2005, 
Cunha2006}. According to these works, a pattern of nearly equally spaced peaks may 
still be found in particular sections of the frequency spectrum, but it is also 
clear that the magnetic field can distort the regular frequency pattern. This 
problem can be resolved by matching of observed frequencies with frequencies 
calculated from models which include the effect of a magnetic field 
\citep{Saio2005}. 
 
The use of the asymptotic relationship or direct frequency matching requires that 
we associate the observed frequencies with the correct value of $\ell$. In roAp 
stars which show rotational modulation, the value of $\ell$ may be inferred from 
the number and relative amplitudes of the rotational multiplets. However, 
application of the method to HD~24712 gives conflicting results \citep{Saio2010}. 
On the other hand, an alternative mode identification cannot reproduce the 
observed rotational multiplet structure in this star.
 
It is quite clear that we are far from a complete understanding of pulsations in 
roAp stars, even though impressive advances have been made both observationally 
and theoretically over the last few years. The varied properties of roAp 
pulsations present a wide range of interesting challenges. Each roAp star seems to 
have its own peculiar characteristics. There were no previously known roAp stars 
in the {\it Kepler} field of view, though several Ap stars were known prior to the 
satellite's launch. It was therefore of great interest to discover that one of 
these stars pulsates in several frequencies in the roAp star range. This star, 
KIC\,8677585 (BD+44\,3063; ILF1+44\,20; JD2000 position: 19:06:28, +44:50:33, $V = 
10.3$) is classified as A5p \citep{Macrae1952}. In this paper we present a 
frequency analysis of this new roAp star and investigate its relationship with 
other roAp and the $\delta$\,Sct and $\gamma$\,Dor variables. 
 
\begin{figure*}  
\centering 
\includegraphics{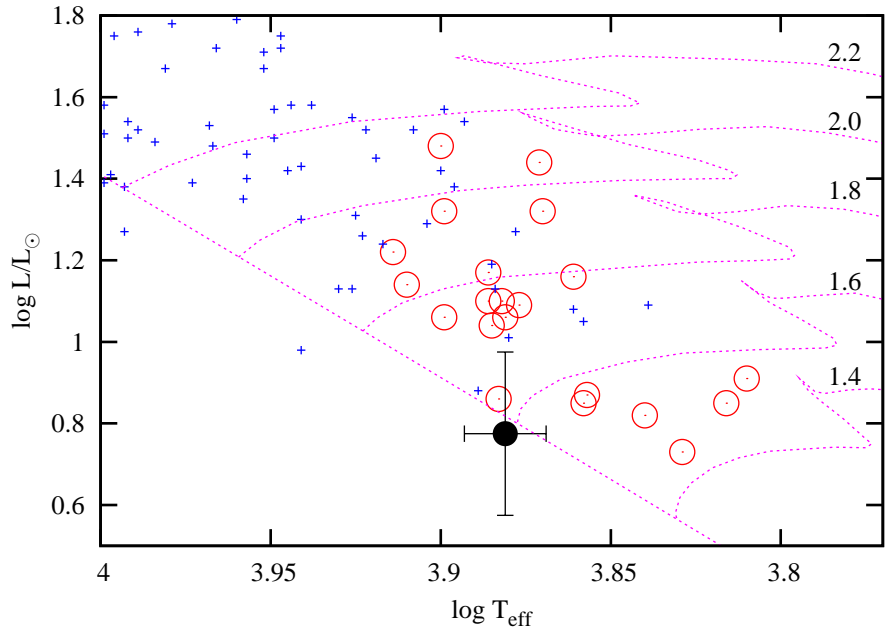}  
\caption{Location of the roAp stars (open circles) and noAp stars (crosses) in  
the theoretical HR diagram. Also shown is the zero-age main sequence and  several 
evolutionary tracks labeled in solar masses. The location of  KIC\,8677585 is 
shown by the filled circle.  The evolutionary tracks were calculated with the
Warsaw-New Jersey code with no convective overshoot.}  
\label{fig:roAp} 
\end{figure*}  
  
\section[]{Observations}

The {\it Kepler} data for KIC\,8677585 consist of 14\,264 nearly continuous 
photometric data points taken during the time interval (truncated Barycentric
Julian date) BJD\,54953.53--54963.25 (9.73\,d) with a mean sampling time of 
59.06\,s (the exposure time is 58.85\,s). The ``uncorrected'' data used 
here consist of aperture photometry in which a few points have been flagged as 
bad for one reason or another and a few more deviate significantly from the mean. 
Removing these points leaves 14\,222 data points that can be analysed. 
KIC\,8677585 has a very small contamination factor of less than 3 percent, so the 
probability of any of the pulsation signal measured coming from a nearby star is 
low.
 
The only ground-based photometric observations available prior to Kepler launch 
for KIC\,8677585 were five-colour photometry using Sloan filters from the {\it 
Kepler Input Catalogue} (KIC). The KIC also contains a number of derived 
quantities including the stellar radius, $R/R_\odot$, the effective temperature, 
$T_{\rm eff}$, and the surface gravity $\log g$. There is no guarantee that these 
values are appropriate, especially for Ap stars. For KIC\,8677585 these are $R = 
1.6$\,R$_\odot$, $T_{\rm eff} = 7400$\,K and $\log g = 4.2$, values that are 
reasonable for a cool Ap star. From these values a luminosity of $\log L/{\rm 
L}_\odot = 0.8$ can be estimated. Frequently, estimates of fundamental parameters 
for Ap stars from photometry are poor because of the strong line blanketing caused 
by overabundances of rare earth elements by orders of magnitude compared to normal 
stars. Nevertheless, for KIC\,8677585 spectroscopic estimates of $T_{\rm  eff}$ 
and $\log g$ are in good agreement with those of the KIC. 

\subsection{Stellar parameters of KIC\,8677585}

One high resolution spectrum of the star was obtained using HERMES (High 
Efficiency and Resolution Mercator Echelle Spectrograph, \cite{Raskin2008}) 
installed at 1.2-m 
Mercator telescope at the Roque de los Muchachos Observatory on La Palma,
Canary Islands (http://www.mercator.iac.es).  The resolving power of HERMES
is about 85\,000. The reduction was done with software developed 
for this instrument. 

From the magnetically null Fe\,\textsc{i} line at 5434.523\,\AA\  we measured the 
projected rotational velocity to be $v \sin i = 4.2 \pm 0.5$\,km\,s$^{-1}$. Lines 
of Nd\,\textsc{iii}, Pr\,\textsc{iii} and Eu\,\textsc{ii} are very strong, as is 
typical for the roAp stars. The equivalent width of the interstellar 
Na\,\textsc{i} D1 line (39\,m\AA) and an empirical calibration by \citet{Munari97} 
shows that the colour excess for the star is less than $E_{B-V}=0.05$.

The Balmer lines profiles are good indicators of effective temperature for roAp 
stars. We have compared observed and synthetic profiles of the H$\alpha$ and 
H$\beta$ lines. The synthetic calculations of Balmer line profiles were done using 
the {\small SYNTH}  code by \citet{Piskunov92} with model atmospheres from the 
NEMO grid \citep{heiteretal02}. By fitting the Balmer profiles we derived 
$T_{\mathrm{eff}} = 7600 \pm 200$\,K.  The Balmer line profiles are not very 
sensitive to $\log g$ at this temperature.  We estimated  $\log g = 4.0 \pm 0.3$ 
from the ionisation equilibrium of Fe\,\textsc{i}  and Fe\,\textsc{ii} lines and 
Cr\,\textsc{i} and Cr\,\textsc{ii} lines, taking into account the indication from 
photometry.

\subsection{Magnetic field}\label{mgfield}

The mean magnetic field modulus $\langle B \rangle$ can be detected in Ap stars 
from high resolution spectra using spectral lines with resolved Zeeman components. 
The line of Fe\,\textsc{ii} at 6149.24\,\AA\ is commonly used for this in cool Ap 
stars. In KIC\,8677585 it shows partial Zeeman splitting. To obtain the mean 
magnetic field modulus we calculated synthetic spectra with the {\small SYNTHMAG} 
code of \citet{Piskunov99} for a range of abundances and magnetic field strengths. 
{The spectral line list was taken from the Vienna Atomic Line Database (VALD, 
\citealt{kupkaetal99}), which includes lines of rare earth elements from the DREAM 
database \citep{biemontetal99}.}  The synthetic spectra were then compared with 
the observations for the best match which yields an estimate of the magnetic field 
modulus of $\langle B \rangle = 3.2 \pm 0.2$\,kG  from line of Fe\,\textsc{ii} 
6149.24\,\AA. We note that this line in KIC\,8677585 is blended in the blue wing, 
probably with the line of Sm\,\textsc{ii} 6149.063\,\AA. 

\section{Position in the HR Diagram}

In Fig.\,\ref{fig:roAp} we show the location of KIC\,8677585 in the 
theoretical HR diagram and compare it with other roAp stars and with noAp stars 
for which temperatures and luminosities are available in the literature. 
We have used the spectroscopic temperature $\log T_{\rm eff} = 3.881 \pm
0.012$ and the $\log L/L_\odot = 0.8 \pm 0.2$ which is the luminosity from
the KIC parameters and assuming a standard deviation of 0.5~mag for the
absolute magnitude.

KIC\,8677585 appears to be on the zero-age main sequence with an effective 
temperature close to the mean value for the roAp stars. Further observations are 
desirable to refine the effective temperature and luminosity. 
 
\begin{figure*} 
\centering 
\includegraphics{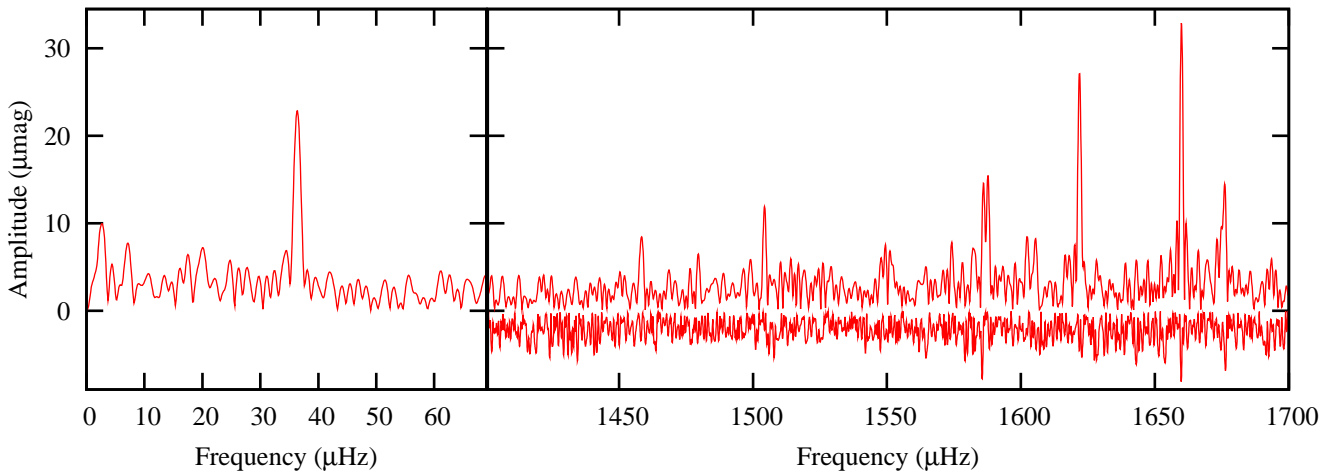} 
\caption{Periodograms of KIC\,8677585. The left panel shows the region
around $f_3 = 36.7~\mu$Hz and the right panel the region showing roAp-type
pulsation.  In this region, a section of the periodogram 
with twice the frequency has been plotted upside down to show the probable 
existence of the first harmonic.}
\label{fig:per} 
\end{figure*}

\begin{center} 
\begin{table} 
 \caption{Frequencies , $f$ ($\mu$Hz), amplitudes, $A$ ($\mu$mag), and phases 
(radians) extracted from {\it Kepler} data of KIC\,8677585 for a model $V = V_0 
+ \sum _n A_n\sin(2\pi f_n (t-t_0) + \phi_n)$, where the epoch of phase zero is 
$t_0$ = BJD\,54950.000.  The last column is the Scargale false alarm
probability (FAP) as calculated by {\it SigSpec} \citep{Reegen2007}.} 
\label{tab:freq}
\begin{tabular}{rrrrr} 
  \hline 
 $N$   & \multicolumn{1}{c}{$f$} &  \multicolumn{1}{c}{$A$}& \multicolumn{1}{c}{$\phi$} & FAP \\ 
 \hline 
   1   &  $1659.919 \pm 0.036$ & $32.9 \pm 1.7$ & $ 0.29 \pm 0.05$ &  0.00  \\ 
   2   &  $1621.856 \pm 0.044$ & $27.1 \pm 1.7$ & $ 0.01 \pm 0.06$ &  0.00  \\ 
   3   &  $  36.359 \pm 0.052$ & $23.1 \pm 1.7$ & $-1.50 \pm 0.08$ &  0.00  \\ 
   4   &  $1587.709 \pm 0.076$ & $15.1 \pm 1.8$ & $ 0.76 \pm 0.12$ &  0.00  \\ 
   5   &  $1676.042 \pm 0.082$ & $14.7 \pm 1.7$ & $ 1.07 \pm 0.12$ &  0.00  \\ 
   6   &  $1586.062 \pm 0.088$ & $14.2 \pm 1.8$ & $ 0.38 \pm 0.13$ &  0.00  \\ 
   7   &  $1504.307 \pm 0.101$ & $11.9 \pm 1.7$ & $ 0.66 \pm 0.15$ &  0.00  \\ 
   8   &  $1674.749 \pm 0.131$ & $\phantom{0}9.1  \pm 1.7$ & $ 1.97 \pm 0.19$ & 0.02 \\ 
   9   &  $   2.611 \pm 0.116$ & $10.0 \pm 1.8$ & $-0.93 \pm 0.17$ &            0.03 \\ 
  10   &  $1458.451 \pm 0.142$ & $\phantom{0}8.4  \pm 1.7$ & $-0.61 \pm 0.21$ & 0.04 \\  
  11   &  $1602.281 \pm 0.137$ & $\phantom{0}8.7  \pm 1.7$ & $ 2.56 \pm 0.20$ & 0.19 \\  
\hline 
\end{tabular} 
\end{table} 
\end{center}

\section{Frequency analysis} 

A periodogram of the light curve shows that KIC\,8677585 varies in two quite 
distinct frequency regions: a low-frequency and a high-frequency domain. In the 
low-frequency domain the only periodicity is $f_0 = 3.141 \pm 0.004$\,d$^{-1}$ 
($36.359 \pm 0.052$\,$\mu$Hz) with an amplitude of $A = 23.1 \pm 1.7$\,$\mu$mag. 
The high-frequency domain contains at least seven components in the range 1450--
1680\,$\mu$Hz. Outside these two regions, the background noise level in the 
periodogram is approximately 5\,$\mu$mag. The significant frequencies are listed 
in Table\,\ref{tab:freq}. In the table we display the false alarm probability
(FAP) \citep{Scargle1982}, showing that there is some doubt about the reality of 
$f_8, f_9, f_{10}$ and especially $f_{11}$.  It should be noted that that the FAP 
is only to be taken as a guideline.   Different approaches to calculate it yield 
different probabilities, which need to be interpreted in the context of their underlying 
definitions and assumptions.  We consider $f_9$ to be almost certainly of instrumental 
origin since {\it Kepler} data have not yet been corrected for long-term drift. 
 
The periodograms in the two regions are shown in Fig.\,\ref{fig:per}. In this 
figure the region around twice the frequency of the main peaks is plotted to show 
the probable detection of the first harmonic of $f_1$ and $f_6$. Harmonics are 
quite common in the light curves of roAp stars, but it seems extraordinary that 
they should be present in the extremely low amplitude variations seen here.

\begin{figure} 
\centering 
\includegraphics{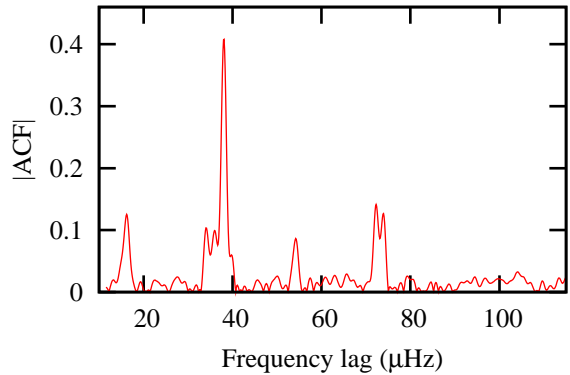} 
\caption{Absolute value of the autocorrelation as a function of frequency lag (in 
$\mu$Hz) for the frequency range 1390--1850\,$\mu$Hz.} 
\label{fig:spacing} 
\end{figure}

Inspection of Fig.\,\ref{fig:per} shows a certain regularity in the frequency 
spacing of the peaks. One can quantify this by calculating the autocorrelation 
function (ACF) of the spectral significance \citep{Reegen2007} within a certain frequency range. 
Fig.\,\ref{fig:spacing} shows the absolute value of the ACF as a function of 
frequency spacing for the frequency range 1390--1850\,$\mu$Hz. There is no change 
when restricting the frequency range even further, but the ACF obviously gets 
worse as the range is expanded. From the figure, the highest peak occurs at a 
frequency lag of 38.1\,$\mu$Hz with smaller peaks at 72.3, 73.9, 16.2 and 
54.2\,$\mu$Hz. From these values the large separation most likely is in the ranges 
[36--39]~$\mu$Hz or [72--76]~$\mu$Hz. 
 
\section{Modelling of KIC\,8677585} 
 
Despite the limitations posed by the lack of reliable stellar parameters, the 
frequencies found in the high-frequency domain do provide sufficient information 
for a first modelling of this star. 
 
As mentioned in the introduction, among other effects, the magnetic field modifies 
the oscillation frequencies. A typical oscillation spectrum of a multiperiodic 
roAp star may still show a pattern of frequencies that resembles that found in a 
non-magnetic star, but often some of the frequencies are displaced in relation to 
the non-magnetic positions. Also, the large separations are usually slightly 
enlarged (by a few $\mu$Hz) in relation to their non-magnetic counterparts. 
 
When the effect of the magnetic field is not taken into account in the models, 
matching of observed and calculated frequency differences are certainly to be 
preferred to matching of individual frequencies themselves, particularly when a 
regular pattern is found. With this in mind, we will use the effective temperature 
and $\log g$ derived from the spectrum of the star, i.e, $T_{\rm eff}=7600\pm 
200$\,K and $\log g = 4.0\pm 0.3$, and the two possible intervals for the observed 
large separation, namely [36--39]~$\mu$Hz and [72--76]~$\mu$Hz, to constrain the 
models. 
 
We considered a grid of possible equilibrium models for the star, obtained using 
the ASTEC stellar evolution code \citep{Christensen2008a}.  We calculated 
evolutionary sequences of models for masses in the range 1.3--2.1\,M$_\odot$, 
in steps of 0.2\,M$_\odot$, setting the initial relative abundance of metals at 
the surface to $Z/X=0.0229$ and the helium abundance to $Y=0.26$. Moreover, we 
considered no core overshooting and a fixed value for the mixing length parameter 
of $\alpha=1.8$. The intention was not to do a detailed modelling of the star, 
varying all possible parameters in a fine grid. That we hope to perform later, 
when significantly longer time-series of Kepler data for this star become 
available. For the moment we are mostly interested is finding how the choice of 
the interval for the large separation influences the basic properties of the star. 
 
\begin{figure} 
\centering 
\includegraphics[scale=0.40]{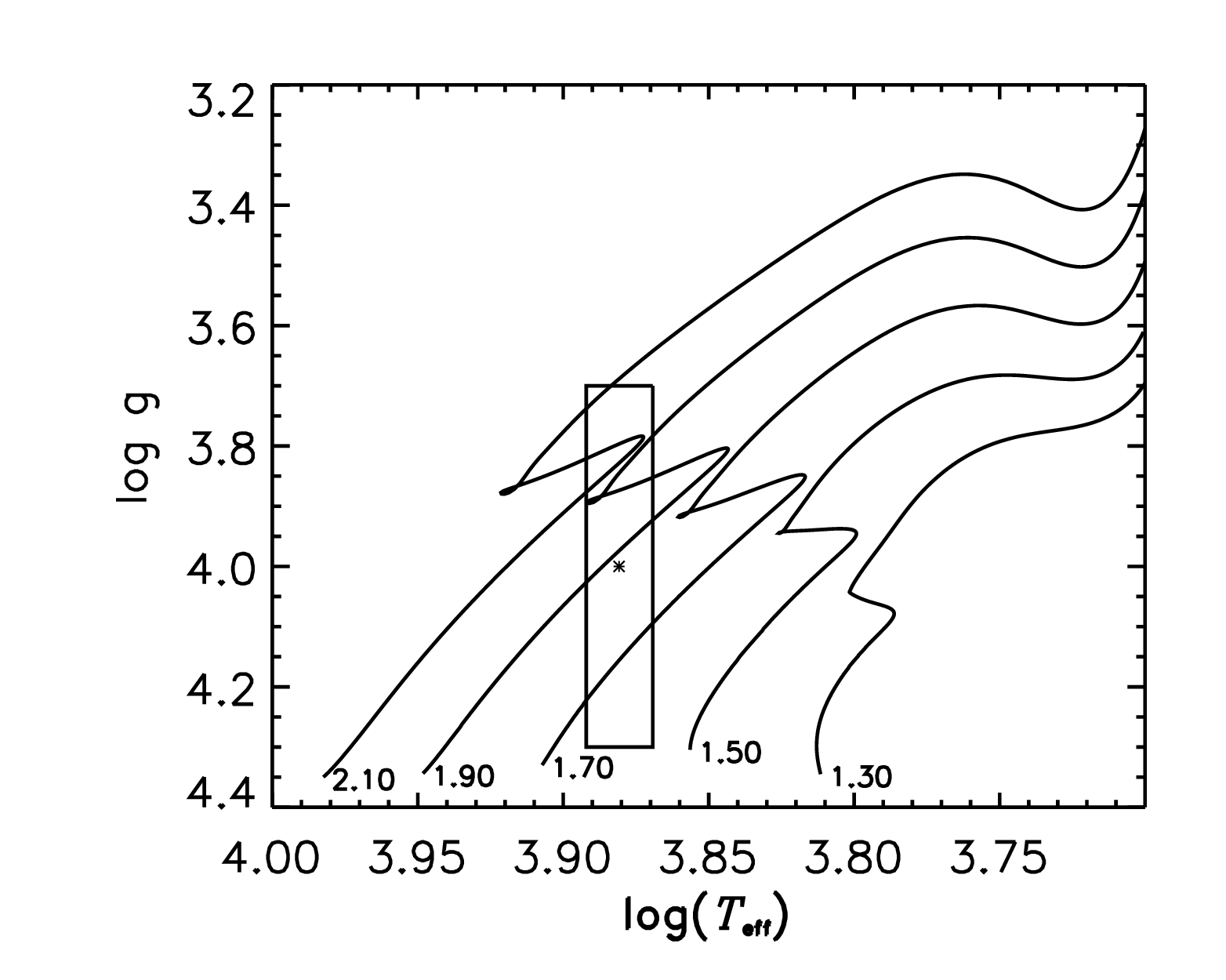} 
\includegraphics[scale=0.40]{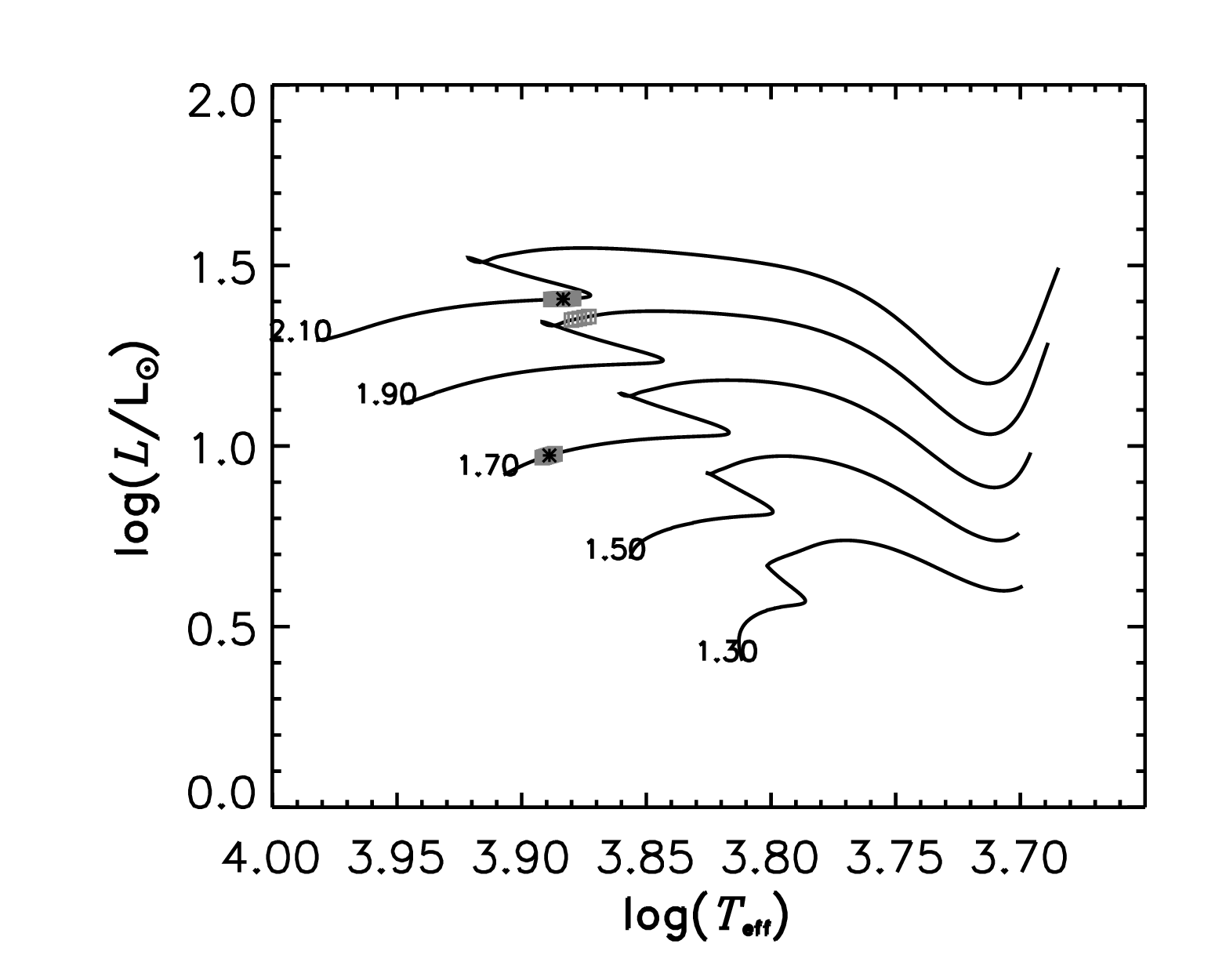} 
 \caption{$\log g$ - $\log T_{\rm eff}$ (upper panel) and $\log L/L_\odot$
- $\log T_{\rm  eff}$ (lower panel) diagrams showing the evolution sequences considered in the 
modelling. The position of KIC\,8677585 is shown in the upper panel by a star, 
along with the 1-$\sigma$ error box. In the lower panel, models with large 
separations within either of the two intervals allowed by the observations are 
shown by grey stars. The properties of the two models marked by black symbols are 
given in Table~\ref{models}. } 
\label{fig:HRmodels} 
\end{figure}

In Fig.~\ref{fig:HRmodels} we show the evolutionary sequences considered here in 
the $\log g - \log T_{\rm eff}$ and $\log L/L_\odot - \log T_{\rm eff}$ diagrams. 
The position of KIC\,8677585, along with the 1$\sigma$ error box is also shown in 
the former. For models within the error box, we then computed the oscillation 
frequencies using the ADIPLS code \citep{Christensen2008b} and, from these, derived 
the corresponding large separations. 
 
\begin{figure} 
\includegraphics[scale=0.37]{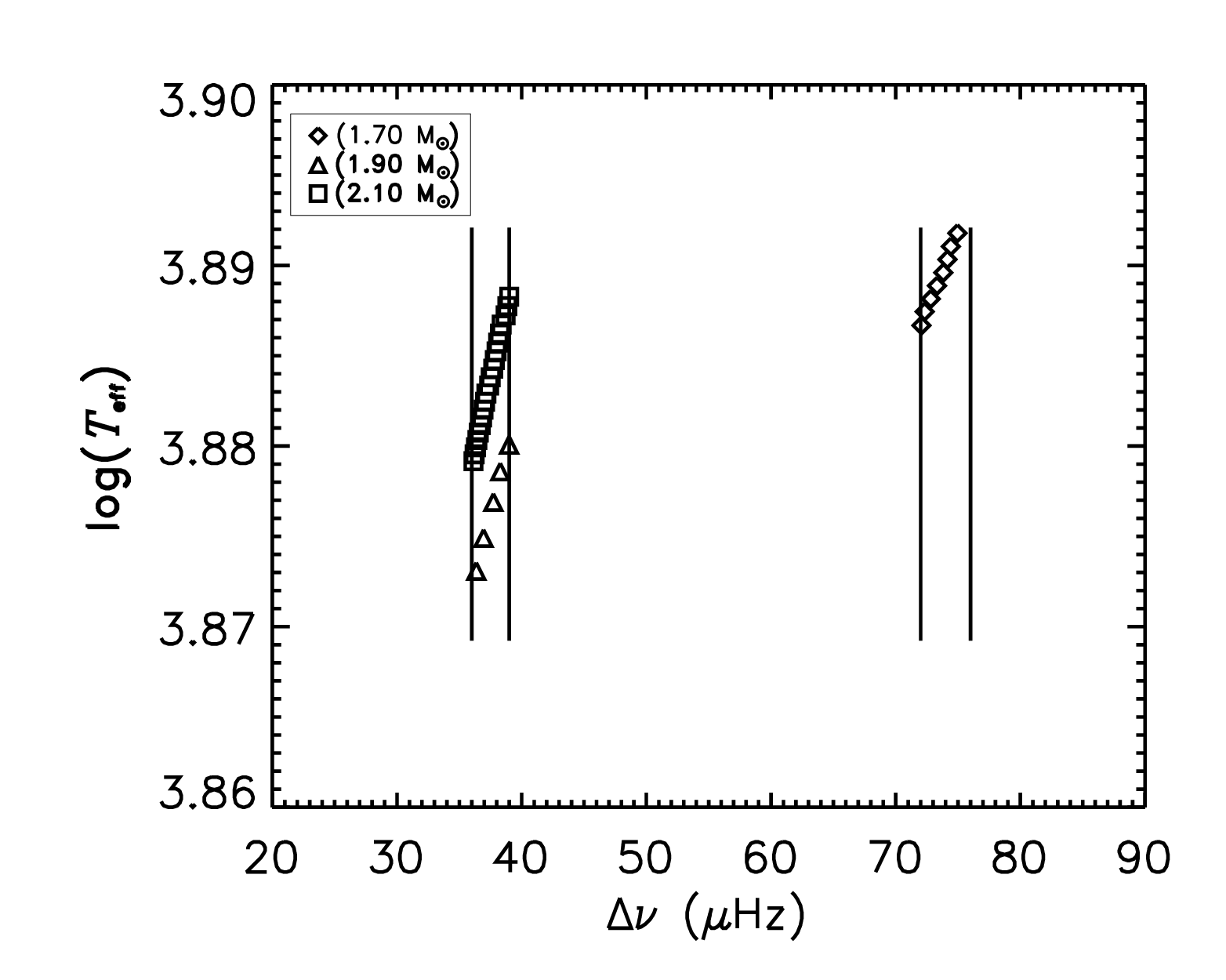} 
\caption{ $T_{\rm eff}$ - $\Delta\nu$ diagram: vertical lines show the limits of 
the possible intervals for the large separation, namely [36--39]~$\mu$Hz and [72--
76]~$\mu$Hz. Different symbols show models of different mass whose large frequency 
separation was found to be within one of the above intervals.}
\label{fig:T-LS} 
\end{figure}

Fig.\,\ref{fig:T-LS} shows the $T_{\rm eff} - \Delta\nu$ diagram along with the 
models whose large separations were found to be within either of the two intervals 
considered. 

In Fig.~\ref{fig:HRmodels} the position of these models in the HR diagram is 
shown, along with corresponding evolutionary tracks.

\begin{center}  
\begin{table}  
 \caption{Data for the best models derived from the the non-magnetic grid.} 
\label{models}  
\begin{tabular}{lrr}             
\hline
                      & Model 1 & Model 2      \\
\hline
$M/M_\odot$           &  2.1     & 1.7         \\
$R/R_\odot$           &  2.89    & 1.71        \\
age(Gyr)              &  0.74628 & 0.59179     \\
$T_{\rm eff}$ (K)     &  7645    & 7742        \\
$L/L_\odot$           &  25.53   & 9.42        \\
$\log~g$              &  3.839   & 4.203       \\
$\Delta\nu$ ($\mu$Hz) &  37.39   & 73.31       \\  
\hline                            
\end{tabular}  
\end{table}  
\end{center}  
 
\begin{figure*} 
\centering 
\includegraphics[scale=0.60]{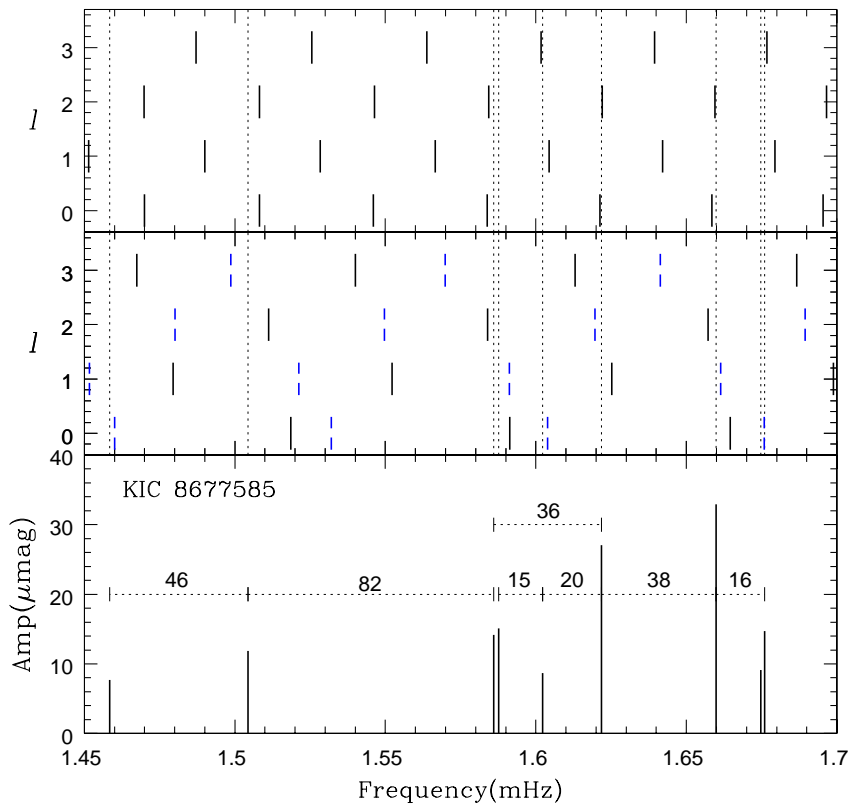} 
\caption{The bottom panel is a schematic depiction of the observed frequencies in 
KIC\,8677585. The numbers along the dotted horizontal lines indicate
frequency differences in $\mu$Hz. Solid vertical lines in the other panels show 
the location of calculated frequencies for $\ell = 0$ to 3 modes in models 1 (top panel) 
and 2 (middle panel) (cf.\ Table \ref{models}). 
The dashed (blue) lines in the middle panel are frequencies calculated including 
the effect of a dipole magnetic field ($B_p=4.2$kG) for a model similar to model~2.}  
\label{schematic} 
\end{figure*} 

\begin{figure*} 
\centering 
\includegraphics[scale=0.60]{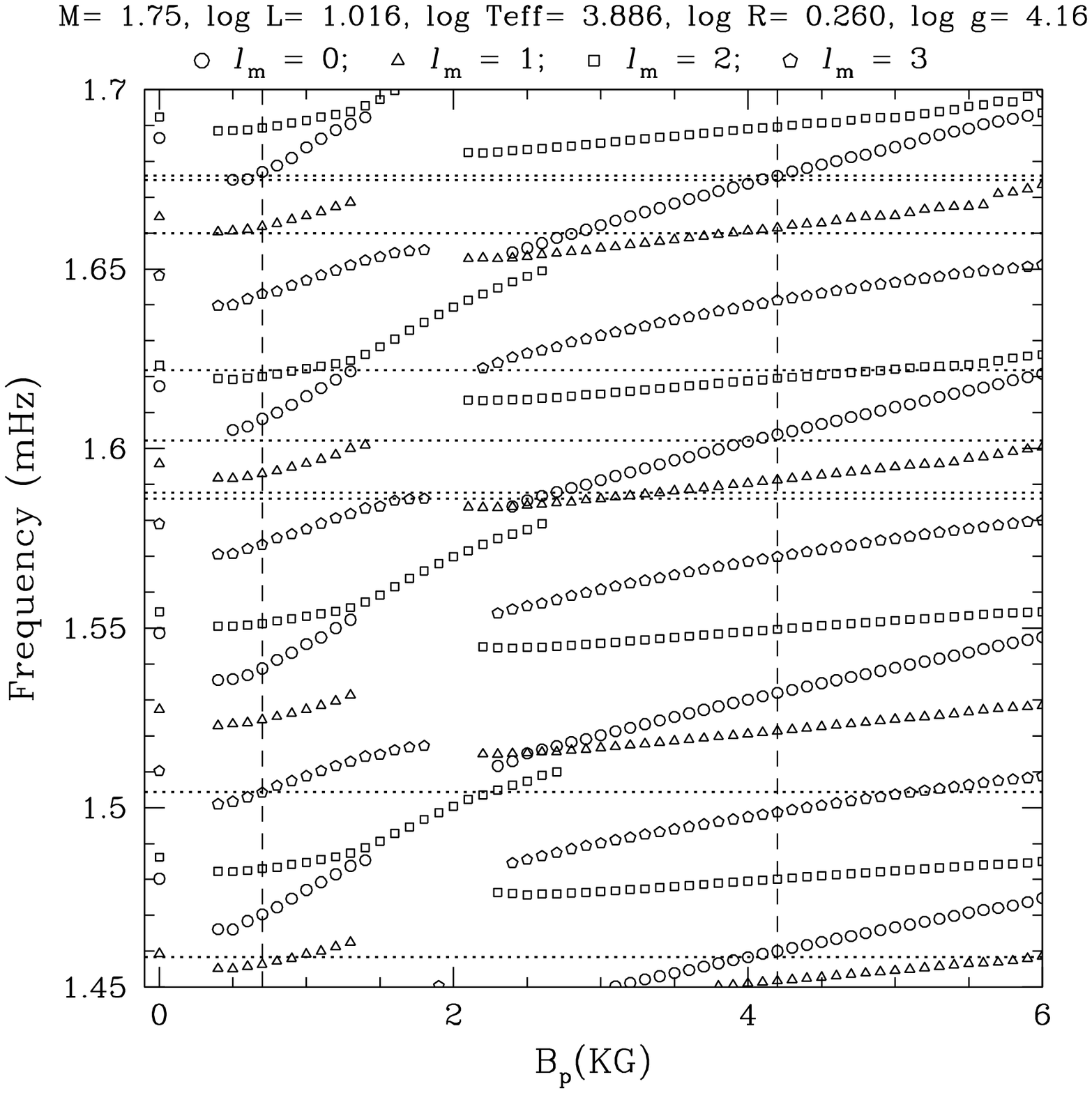} 
\caption{Oscillation frequency versus magnetic field strength at poles $B_p$.
Horizontal dotted lines indicate observed frequencies of KIC\,8677585, while vertical
dashed lines indicate the magnetic field strengths which yield frequencies roughly agree
with the observed ones.}
\label{freqbp} 
\end{figure*} 
 
The oscillation frequencies of the two best models of our grid are shown in the 
upper and middle panels of Fig.~\ref{schematic}. A schematic amplitude spectrum of 
KIC\,8677585 in the high-frequency domain, with an indication of the frequency 
separations between different peaks, is shown for comparison in the lowest panel. 
The best models correspond to those that minimize a $\chi^2$ function taking into 
account both the non-seismic data and the large separation. For the latter we 
considered the two intervals determined from the observations, hence the existence 
of two best models. The properties of these two models are presented in 
Table~\ref{models}. 
 
From the figure one could naively say that models with the smaller option for the 
large separation reproduce best the seismic data. However, that is a dangerous 
statement to make for a number of reasons: First, the comparison of individual 
frequencies, between models and observations, is known to be risky due to possible 
systematic effects that may result from the incorrect modelling of the surface 
layers of stars; secondly, no detailed exploitation of the parameter space was 
considered in the modelling, i.e, the step considered in stellar mass was large 
and no variation of chemical composition or convective overshoot was considered in 
the present grid; thirdly, the effect of the magnetic field on the oscillations 
was not taken into account.

An example of the importance of taking these effect into account is shown in the 
middle panel of Fig.~\ref{schematic} where the dashed lines show 
the model frequencies obtained including the effect of 
a dipole magnetic field with a polar strength of $B_p=4.2$~kG.
Those frequencies have been obtained by the method of non-adiabatic analysis described in 
\citet{Saio2005}, where the oscillations are assumed to be axisymmetric
to the magnetic axis. 
The latitudinal dependence of an axisymmetric oscillation mode is 
represented by a sum of components proportional to spherical harmonics $Y_\ell^{m=0}$.
The latitudinal degree of each mode is identified by $l_m$ which is the $\ell$ value for
the component having a largest kinetic energy.    
The parameters of unperturbed model (shown in Fig.~\ref{freqbp}) are similar to those
of model 2 in Table~\ref{models}, but a slightly larger mass of $1.75M_\odot$ is adopted because
a composition of $(X,Z)=(0.7,0.02)$ is used.  

Although the large separation for a given latitudinal degree hardly changes,
the relative frequencies between different degrees are affected by the presence
of a magnetic field.  This comes from the fact that modes of different degree are 
affected differently by the magnetic field, as seen in Fig.~\ref{freqbp} where 
oscillation frequencies are shown as a function of $B_p$.
In this model, calculated frequencies agree with observed ones at $B_p\approx 4.2$kG and
$\approx 0.7$kG as indicated by vertical lines in Fig.~\ref{freqbp}.
Model frequencies are similar at the two different values of $B_p$ because the magnetic effect
changes cyclically as first found by \citet{Cunha2000}. The former value of
$B_p$ is consistent with the mean modulus $\langle B\rangle = 3.2$kG from 
the spectroscopic analysis (\S~\ref{mgfield}).
We note that, in addition to the frequencies listed in Table~\ref{tab:freq}, the periodogram 
(Fig.~\ref{fig:per}) shows minor peaks at $\approx 1.55$mHz and $\approx 1.69$mHz 
(and possibly at $\approx 1.48$mHz) which correspond to frequencies of $l_m=2$ modes.  
A future long-term observation will clarify the reality of these frequencies. 
Then, we will be able to better constrain the stellar parameters, and to examine the
theory for the oscillations of magnetic stars.   

Although all the frequencies shown in Fig.~\ref{freqbp} are below the critical acoustic
frequency, no high-frequency modes in this model are excited. 
(We note that for low-mass models ($M \la 1.6 M_\odot$) high-frequency modes in the
observed frequency range are excited by the kappa-mechanism in the hydrogen ionization zone.)
Since the stability of high-order p modes seems sensitive to the treatment of 
the optically thin layers, we need further investigation on the excitation of the
high-order p modes.
 
\begin{figure*} 
\centering 
\includegraphics{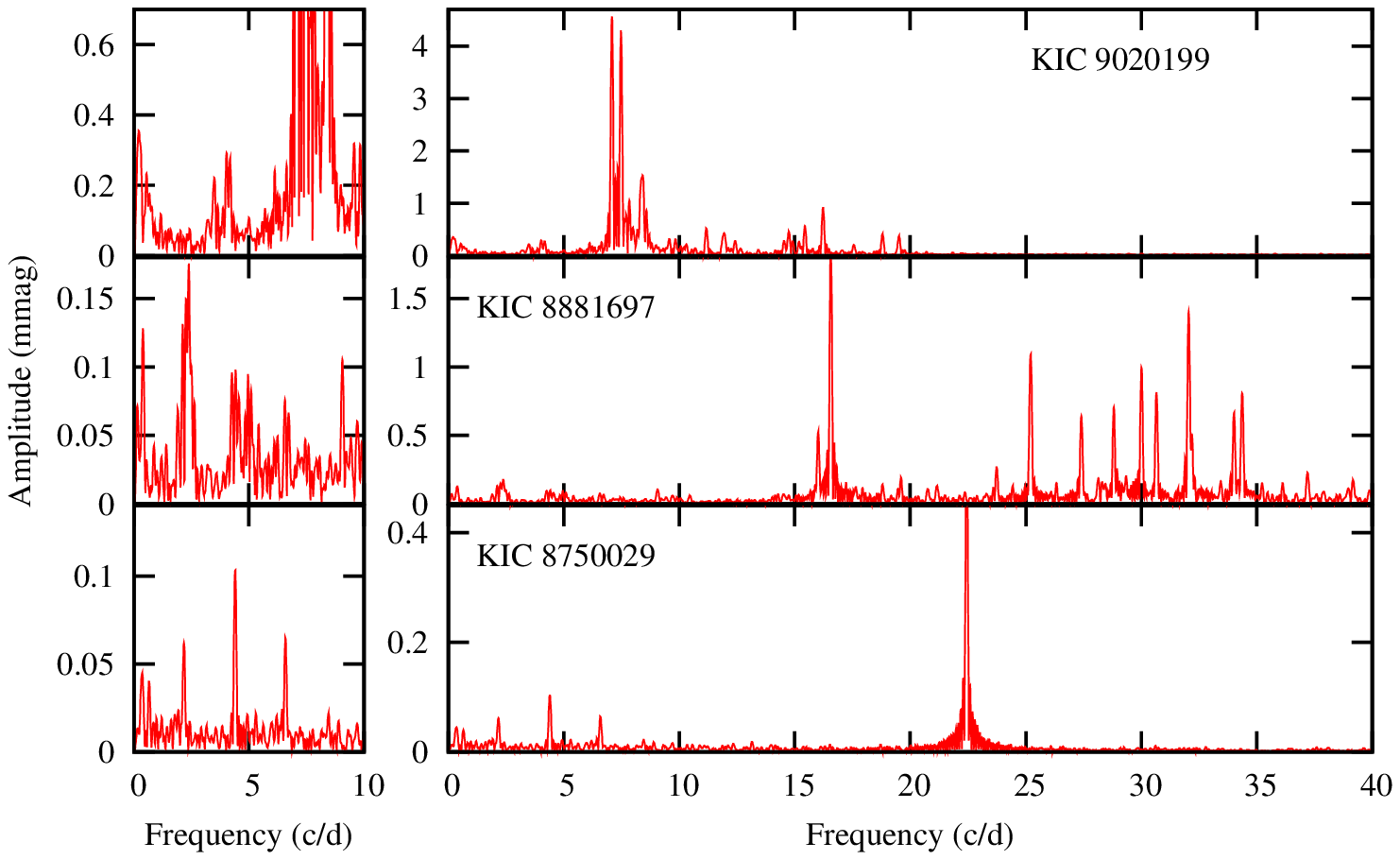} 
 \caption{Periodograms Ap stars observed with {\it Kepler} showing both 
$\delta$\,Sct variations and multiperiodic $\gamma$\,Dor variations. The 
frequencies are is in cycles d$^{-1}$ and the amplitudes in mmag. The left panel 
shows the low frequency region on an expanded amplitude scale.} 
\label{fig:dsct} 
\end{figure*}

\section{Discussion} 
 
The frequency of the pulsations seen in KIC\,8677585 are typical of roAp stars, 
but the amplitudes are an order of magnitude below any that have been detected 
from ground-based observations. The outstanding feature of this star, however, is 
the presence of the low frequency at $f_3 = 3.142$\,d$^{-1}$ which has never been 
seen in roAp stars. A simple calculation shows that the critical rotational 
frequency for roAp stars is typically in the range 2 - 5\,d$^{-1}$. If $f_3$ is 
due to rotation, the star must be rotating at close to the critical rate. All 
known roAp stars rotate slowly, and the measured $v \sin i = 4.2 \pm 
0.5$\,km\,s$^{-1}$ for this star effectively rules out the possibility that $f_3$ 
could be the rotation frequency. A secondary body orbiting just above the 
photosphere could have an orbital frequency with this value. However, the light 
curve at $f_3$ is a pure sinusoid, not an eclipse, but it could conceivably be a 
tidal distortion. 
 
The limiting factor for observations of faint sources is set by source confusion, 
rather than the photometric accuracy computed for isolated sources.  An estimate 
of the crowding metric calculated from the point spread function of surrounding 
objects is provided for most sources in the KIC.  For KIC\,8677585 the
contamination coefficient is 0.029.  It is therefore not impossible that $f_3$ 
may be due to the small contribution from a (faint) neighbouring star.  To eliminate 
this possibility requires extensive photometry with a very small aperture with a 
precision comparable to that of {\it Kepler}. For $f_3$ to arise in the much 
fainter contaminating star would require that this be the only frequency of 
variation of that star, which again is unlikely. A spectrum of the speculative 
faint neighbour is required for better understanding of the probability.
 
\begin{table}  
\begin{center}  
\caption{Ap stars observed by {\it Kepler}, The last column is a  
classification based on the visibility of high ($\delta$\,Sct) and low 
($\gamma$\,Dor) frequencies.}  
\label{tab:Ap} 
\vspace{2mm}  
\begin{tabular}{ r@{\hspace{2mm}} l@{\hspace{2mm}}  
l@{\hspace{2mm}} r@{\hspace{2mm}} l@{\hspace{2mm}}}  
\hline  
      KIC &   Name            & SpType     & Mag & Classification \\  
\hline  
\\  
 8677585 & ILF1+44 20         & A5p        & 10.27 &  roAp/$\gamma$\,Dor  \\  
 8750029 & ILF1+44 257        & A5p?       &  9.66 &  
$\delta$\,Sct/$\gamma$\,Dor \\  
 8881697 & ILF1+44 259        & A5p        & 10.57 & $\delta$\,Sct \\  
 9020199 & HD 182895          & F0p        &  8.86 &  
$\delta$\,Sct/$\gamma$\,Dor \\  
 9147002 & HD 180239          & A2p        &  9.94 & Constant \\  
 9216367 & ILF1+45 298        & A2p:       & 12.12 & $\gamma$\,Dor? \\  
 9640285 & ILF1+46 40         & F5p        & 11.77 & Constant? \\  
\\  
\hline  
\end{tabular}  
\end{center}  
\end{table}  
 
In general, low-frequency pulsations in Ap stars are thought to be unlikely. The 
settling and diffusion which give rise to the chemical inhomogeneities in these 
stars drain out He from the He\,\textsc{ii} driving zone, rendering them 
pulsationally stable. In addition, damping of pulsations is expected in these 
stars due to magnetic slow wave leakage if the magnetic field strength is greater 
than about 1\,kG. Prior to the {\it Kepler} observations, the only Ap star known 
to pulsate in low frequencies is HD~21190 (F2III SrEuSi). In this star there is at 
least one mode with a frequency of 6.68~d$^{-1}$ which is probably a radial mode 
\citep{Koen2001}. The star is multiperiodic, but other frequencies could not be 
extracted from the photometric data. High-time resolution spectroscopic 
observations of HD~21190 show moving bumps in the cores of spectral lines, 
indicating the presence of high-degree nonradial pulsations \citep{Gonzalez2008}. 
HD~21190 is the most evolved Ap star known; its unique stage of evolution may 
offer clues as to why it pulsates. 
 
{\it Kepler} observations of other Ap stars indicate that pulsations of low radial 
order may be more common than previously thought. Table\,\ref{tab:Ap} lists 
information for stars in the {\it Kepler} field of view which have been classified 
as Ap. However, these classifications have never been confirmed and they may turn 
out to be in error. Ap and Am classifications are easily confused at 
classification dispersions. In at least three of these stars $\delta$\,Sct 
pulsations are clearly present. In two of the stars, multiperiodic $\gamma$\,Dor 
pulsations are also visible (Fig.\,\ref{fig:dsct}). 
 
The clear presence of $\gamma$\,Dor pulsations in two presumed Ap stars is most 
surprising. The multiperiodicity renders any explanation in terms of orbital or 
tidal interaction invalid. This has a bearing on the $f_3$ mode in KIC\,8677585. 
Clearly if the presumed Ap stars with low-frequency modes are indeed Ap stars, it 
means that the low-frequency $f_3$ mode in KIC\,8677585 is not unique to the star 
and requires an explanation in the context of Ap stars as a whole rather than 
specific to this star. 
  
As already mentioned, models suggest that pulsations in Ap stars are not excited 
because of draining of He from the driving zone and damping by magnetic slow wave 
leakage. The presence of a low frequency in KIC\,8677585 and other Ap stars 
contradicts these ideas. Confirmation of the classifications of the stars in 
Table\,\ref{tab:Ap} are, however, required to cement this conclusion. 
  
\section*{Acknowledgments}  

The authors wish to thank the {\it Kepler} team for their generousity in
allowing the data to be released to the Kepler Asteroseismic Science Consortium 
(KASC) ahead of public release and for their outstanding efforts which have
made these results possible.  Funding for the {\it Kepler} mission is provided 
by NASA's Science Mission Directorate.  We particularly thank Ron Gilliland and 
Hans Kjeldsen for their tireless work on behalf of KASC.

LAB wishes to thank the South African Astronomical Observatory for financial 
support.

This work was partially supported by the project PTDC/CTE-AST/098754/2008
and the grant SFRH / BD / 41213 / 2007 funded by FCT/MCTES, Portugal.
MC is supported by a Ci\^encia 2007 contract, funded by FCT/MCTES(Portugal)
and POPH/FSE (EC).

MG has received financial support through D. Guenther's NSERC grant.

\bibliographystyle{mn2e} 
\bibliography{roap1}

\begin{thebibliography}{29}
\expandafter\ifx\csname natexlab\endcsname\relax\def\natexlab#1{#1}\fi

\bibitem[{{Balmforth} {et~al.}(2001){Balmforth}, {Cunha}, {Dolez}, {Gough}, \&
  {Vauclair}}]{Balmforth2001}
{Balmforth} N.~J., {Cunha} M.~S., {Dolez} N., {Gough} D.~O., {Vauclair} S.,
  2001, \mnras, 323, 362

\bibitem[{{Bi{\'e}mont} {et~al.}(1999){Bi{\'e}mont}, {Palmeri}, \&
  {Quinet}}]{biemontetal99}
{Bi{\'e}mont} E., {Palmeri} P., {Quinet} P., 1999, \apss, 269, 635

\bibitem[{{Bigot} \& {Dziembowski}(2002)}]{bigot2002}
{Bigot} L., {Dziembowski} W.~A., 2002, \aap, 391, 235

\bibitem[{{Christensen-Dalsgaard}(2008{\natexlab{a}})}]{Christensen2008b}
{Christensen-Dalsgaard} J., 2008{\natexlab{a}}, \apss, 316, 113

\bibitem[{{Christensen-Dalsgaard}(2008{\natexlab{b}})}]{Christensen2008a}
---, 2008{\natexlab{b}}, \apss, 316, 13

\bibitem[{{Cunha}(2002)}]{Cunha2002}
{Cunha} M.~S., 2002, \mnras, 333, 47

\bibitem[{{Cunha}(2006)}]{Cunha2006}
---, 2006, \mnras, 365, 153

\bibitem[{{Cunha}(2007)}]{cunha2007}
---, 2007, Communications in Asteroseismology, 150, 48

\bibitem[{{Cunha} \& {Gough}(2000)}]{Cunha2000}
{Cunha} M.~S., {Gough} D., 2000, \mnras, 319, 1020

\bibitem[{{Gilliland} {et~al.}(2010){Gilliland}, {Brown},
  {Christensen-Dalsgaard}, {Kjeldsen}, {Aerts}, {Appourchaux}, {Basu},
  {Bedding}, {Chaplin}, {Cunha}, {De Cat}, {De Ridder}, {Guzik}, {Handler},
  {Kawaler}, {Kiss}, {Kolenberg}, {Kurtz}, {Metcalfe}, {Monteiro}, {Szab{\'o}},
  {Arentoft}, {Balona}, {Debosscher}, {Elsworth}, {Quirion}, {Stello},
  {Su{\'a}rez}, {Borucki}, {Jenkins}, {Koch}, {Kondo}, {Latham}, {Rowe}, \&
  {Steffen}}]{Gilliland2010}
{Gilliland} R.~L., {Brown} T.~M., {Christensen-Dalsgaard} J., {Kjeldsen} H.,
  {Aerts} C., {Appourchaux} T., {Basu} S., {Bedding} T.~R., {Chaplin} W.~J.,
  {Cunha} M.~S., {De Cat} P., {De Ridder} J., {Guzik} J.~A., {Handler} G.,
  {Kawaler} S., {Kiss} L., {Kolenberg} K., {Kurtz} D.~W., {Metcalfe} T.~S.,
  {Monteiro} M.~J.~P.~F.~G., {Szab{\'o}} R., {Arentoft} T., {Balona} L.,
  {Debosscher} J., {Elsworth} Y.~P., {Quirion} P., {Stello} D., {Su{\'a}rez}
  J.~C., {Borucki} W.~J., {Jenkins} J.~M., {Koch} D., {Kondo} Y., {Latham}
  D.~W., {Rowe} J.~F., {Steffen} J.~H., 2010, \pasp, 122, 131

\bibitem[{{Gonz{\'a}lez} {et~al.}(2008){Gonz{\'a}lez}, {Hubrig}, {Kurtz},
  {Elkin}, \& {Savanov}}]{Gonzalez2008}
{Gonz{\'a}lez} J.~F., {Hubrig} S., {Kurtz} D.~W., {Elkin} V., {Savanov} I.,
  2008, \mnras, 384, 1140

\bibitem[{{Gough}(2005)}]{gough2005}
{Gough} D., 2005, The Roger Tayler Memorial Lectures 1999-2003: Special issue
  of Astronomy and Geophysics, 16

\bibitem[{{Heiter} {et~al.}(2002){Heiter}, {Kupka}, {van't Veer-Menneret},
  {Barban}, {Weiss}, {Goupil}, {Schmidt}, {Katz}, \& {Garrido}}]{heiteretal02}
{Heiter} U., {Kupka} F., {van't Veer-Menneret} C., {Barban} C., {Weiss} W.~W.,
  {Goupil} M., {Schmidt} W., {Katz} D., {Garrido} R., 2002, \aap, 392, 619

\bibitem[{{Koch} {et~al.}(2010){Koch}, {Borucki}, {Basri}, {Batalha}, {Brown},
  {Caldwell}, {Christensen-Dalsgaard}, {Cochran}, {DeVore}, {Dunham},
  {Gautier}, {Geary}, {Gilliland}, {Gould}, {Jenkins}, {Kondo}, {Latham},
  {Lissauer}, {Marcy}, {Monet}, {Sasselov}, {Boss}, {Brownlee}, {Caldwell},
  {Dupree}, {Howell}, {Kjeldsen}, {Meibom}, {Morrison}, {Owen}, {Reitsema},
  {Tarter}, {Bryson}, {Dotson}, {Gazis}, {Haas}, {Kolodziejczak}, {Rowe}, {Van
  Cleve}, {Allen}, {Chandrasekaran}, {Clarke}, {Li}, {Quintana}, {Tenenbaum},
  {Twicken}, \& {Wu}}]{Koch2010}
{Koch} D.~G., {Borucki} W.~J., {Basri} G., {Batalha} N.~M., {Brown} T.~M.,
  {Caldwell} D., {Christensen-Dalsgaard} J., {Cochran} W.~D., {DeVore} E.,
  {Dunham} E.~W., {Gautier} III T.~N., {Geary} J.~C., {Gilliland} R.~L.,
  {Gould} A., {Jenkins} J., {Kondo} Y., {Latham} D.~W., {Lissauer} J.~J.,
  {Marcy} G., {Monet} D., {Sasselov} D., {Boss} A., {Brownlee} D., {Caldwell}
  J., {Dupree} A.~K., {Howell} S.~B., {Kjeldsen} H., {Meibom} S., {Morrison}
  D., {Owen} T., {Reitsema} H., {Tarter} J., {Bryson} S.~T., {Dotson} J.~L.,
  {Gazis} P., {Haas} M.~R., {Kolodziejczak} J., {Rowe} J.~F., {Van Cleve}
  J.~E., {Allen} C., {Chandrasekaran} H., {Clarke} B.~D., {Li} J., {Quintana}
  E.~V., {Tenenbaum} P., {Twicken} J.~D., {Wu} H., 2010, ArXiv e-prints

\bibitem[{{Koen} {et~al.}(2001){Koen}, {Kurtz}, {Gray}, {Kilkenny}, {Handler},
  {Van Wyk}, {Marang}, \& {Winkler}}]{Koen2001}
{Koen} C., {Kurtz} D.~W., {Gray} R.~O., {Kilkenny} D., {Handler} G., {Van Wyk}
  F., {Marang} F., {Winkler} H., 2001, \mnras, 326, 387

\bibitem[{{Kupka} {et~al.}(1999){Kupka}, {Piskunov}, {Ryabchikova}, {Stempels},
  \& {Weiss}}]{kupkaetal99}
{Kupka} F., {Piskunov} N., {Ryabchikova} T.~A., {Stempels} H.~C., {Weiss}
  W.~W., 1999, \aaps, 138, 119

\bibitem[{{Kurtz}(1982)}]{Kurtz1982}
{Kurtz} D.~W., 1982, \mnras, 200, 807

\bibitem[{{Macrae}(1952)}]{Macrae1952}
{Macrae} D.~A., 1952, \apj, 116, 592

\bibitem[{{Munari} \& {Zwitter}(1997)}]{Munari97}
{Munari} U., {Zwitter} T., 1997, \aap, 318, 269

\bibitem[{{Piskunov}(1999)}]{Piskunov99}
{Piskunov} N., 1999, in Astrophysics and Space Science Library, Vol. 243,
  Polarization, {K.~N.~Nagendra \& J.~O.~Stenflo}, ed., pp. 515--525

\bibitem[{{Piskunov}(1992)}]{Piskunov92}
{Piskunov} N.~E., 1992, in Stellar Magnetism, {Y.~V.~Glagolevskij \&
  I.~I.~Romanyuk}, ed., p.~92

\bibitem[{{Raskin} \& {Van Winckel}(2008)}]{Raskin2008}
{Raskin} G., {Van Winckel} H., 2008, in Presented at the Society of
  Photo-Optical Instrumentation Engineers (SPIE) Conference, Vol. 7014, Society
  of Photo-Optical Instrumentation Engineers (SPIE) Conference Series

\bibitem[{{Reegen}(2007)}]{Reegen2007}
{Reegen} P., 2007, \aap, 467, 1353

\bibitem[{{Saio}(2005)}]{Saio2005}
{Saio} H., 2005, \mnras, 360, 1022

\bibitem[{{Saio} \& {Gautschy}(2004)}]{Saio2004}
{Saio} H., {Gautschy} A., 2004, \mnras, 350, 485

\bibitem[{{Saio} {et~al.}(2010){Saio}, {Ryabchikova}, \& {Sachkov}}]{Saio2010}
{Saio} H., {Ryabchikova} T., {Sachkov} M., 2010, \mnras, 403, 1729

\bibitem[{{Scargle}(1982)}]{Scargle1982}
{Scargle} J.~D., 1982, \apj, 263, 835

\bibitem[{{Tassoul}(1980)}]{Tassoul1980}
{Tassoul} M., 1980, \apjs, 43, 469

\bibitem[{{Th{\'e}ado} {et~al.}(2009){Th{\'e}ado}, {Dupret}, {Noels}, \&
  {Ferguson}}]{Theado2009}
{Th{\'e}ado} S., {Dupret} M., {Noels} A., {Ferguson} J.~W., 2009, \aap, 493,
  159

\end{thebibliography}

\label{lastpage}  
  
\end{document}